\documentclass[manuscript,sigconf]{acmart}
\AtBeginDocument{%
  \providecommand\BibTeX{{%
    \normalfont B\kern-0.5em{\scshape i\kern-0.25em b}\kern-0.8em\TeX}}}

\usepackage{tabularx}
\usepackage{soul}
\usepackage{cleveref}
\usepackage{setspace}
\makeatletter \AtBeginDocument{\let\hl\@firstofone} \makeatother

\copyrightyear{2024} 
\acmYear{2024} 
\setcopyright{rightsretained} 
\acmConference[CHI '24]{Proceedings of the CHI Conference on Human Factors in Computing Systems}{May 11--16, 2024}{Honolulu, HI, USA}
\acmBooktitle{Proceedings of the CHI Conference on Human Factors in Computing Systems (CHI '24), May 11--16, 2024, Honolulu, HI, USA}
\acmDOI{10.1145/3613904.3641994}
\acmISBN{979-8-4007-0330-0/24/05}

\begin{document}

\title[Charting Ethical Tensions in Multispecies Technology Research]{Charting Ethical Tensions in Multispecies Technology Research through Beneficiary-Epistemology Space}

\author{Steve Benford}
\affiliation{%
  \institution{University of Nottingham}
  \city{Nottingham}
  \country{United Kingdom}}
\email{steve.benford@nottingham.ac.uk}

\author{Clara Mancini}
\affiliation{%
  \institution{The Open University}
  \city{Milton Keynes}
  \country{United Kingdom}}
\email{clara.mancini@open.ac.uk}

\author{Alan Chamberlain}
\affiliation{%
  \institution{University of Nottingham}
  \city{Nottingham}
  \country{United Kingdom}}
\email{alan.chamberlain@nottingham.ac.uk}

\author{Eike Schneiders}
\affiliation{%
  \institution{University of Nottingham}
  \city{Nottingham}
  \country{United Kingdom}}
\email{eike.schneiders@nottingham.ac.uk}

\author{Simon Castle-Green}
\affiliation{%
  \institution{University of Nottingham}
  \city{Nottingham}
  \country{United Kingdom}}
\email{simon.castle-green@nottingham.ac.uk}

\author{Joel Fischer}
\affiliation{%
  \institution{University of Nottingham}
  \city{Nottingham}
  \country{United Kingdom}}
\email{joel.fischer@nottingham.ac.uk}

\author{Ayse Kucukyilmaz}
\affiliation{%
  \institution{University of Nottingham}
  \city{Nottingham}
  \country{United Kingdom}}
\email{ayse.kucukyilmaz@nottingham.ac.uk}

\author{Guido Salimbeni}
\affiliation{%
  \institution{University of Nottingham}
  \city{Nottingham}
  \country{United Kingdom}}
\email{guido.salimbeni@nottingham.ac.uk}

\author{Victor Ngo}
\affiliation{%
  \institution{University of Nottingham}
  \city{Nottingham}
  \country{United Kingdom}}
\email{victor.ngo@nottingham.ac.uk}

\author{Pepita Barnard}
\affiliation{%
  \institution{University of Nottingham}
  \city{Nottingham}
  \country{United Kingdom}}
\email{pepita.barnard@nottingham.ac.uk}

\author{Matt Adams}
\author{Nick Tandavanitj}
\author{Ju Row Farr}
\affiliation{%
  \institution{Blast Theory}
  \city{Brighton}
  \country{United Kingdom}}
\email{{matt,nick,ju}@blasttheory.co.uk}



\renewcommand{\shortauthors}{Benford et al.}
\begin{abstract}
    While ethical challenges are widely discussed in HCI, far less is reported about the ethical processes that researchers routinely navigate. We reflect on a multispecies project that negotiated an especially complex ethical approval process. Cat Royale was an artist-led exploration of creating an artwork to engage audiences in exploring trust in autonomous systems. The artwork took the form of a robot that played with three cats. Gaining ethical approval required an extensive dialogue with three Institutional Review Boards (IRBs) covering computer science, veterinary science and animal welfare, raising tensions around the welfare of the cats, perceived benefits and appropriate methods, and reputational risk to the University. To reveal these tensions we introduce beneficiary-epistemology space, that makes explicit who benefits from research (humans or animals) and underlying epistemologies. Positioning projects and IRBs in this space can help clarify tensions and highlight opportunities to recruit additional expertise. 
\end{abstract}



\maketitle

\section{Introduction}
Although many of us routinely engage with ethical review and Institutional Review Boards (IRBs), there is little reflection on ethical review processes within HCI. This is somewhat surprising given that (i) we are a broad interdisciplinary field drawing together different perspectives that may be in tension; (ii) we increasingly focus on vulnerable communities such as children, people with disabilities and marginalised groups, who may require special ethical consideration; and (iii) we are a relatively young community, typically without our own dedicated IRBs, and so must often seek ethical approval from others. It is a particularly significant omission as HCI turns its attention towards animals as part of multispecies interactions, as historically these stakeholders have proved to be especially vulnerable in terms of their relationship to research and so are now often the subject of strict ethical review processes. 

There has of course been some reflection on ethical review processes within HCI. Brown at al.’s five ethical provocations encourage us to reappraise established processes inherited from medical ethics~\cite{Brown:2016:FiveProvo}, while Munteanu et al.’s reflections on seeking ethical approval for projects involving vulnerable communities advocate for iterative dialogue with IRBs~\cite{Munteanu:2015:SitEthic}. As we discuss further below, two specific areas in which ethical processes have been discussed in some depth are artist-led HCI, where artists undertake exploratory and sometimes provocative projects that transgress accepted norms~\cite{Benford:2013:PerfLed}, and, of course, Animal Computer-Interaction (ACI), where complex challenges arise concerning animals’ consent and autonomy~\cite{Mancini2022relevance}.  

This paper reflects on how one particular project negotiated ethical approval. The project was especially challenging because it combined artist-led HCI with ACI. 
\textit{Cat Royale} was a digital artwork, developed by professional artists Blast Theory in partnership with HCI researchers with the overall aim of exploring the trustworthiness of autonomous systems. \hl{This was to be achieved by striving to create a so-called ‘utopia’ for a family of three cats, a purposefully designed living space at the centre of which a robot would try to enrich their lives by playing with them. From the outset, this somewhat ambiguous artistic framing was intended to provoke audiences to reflect on the role of utopian and dystopian visions of autonomous systems (both  for the cats and by extension for people) while also delivering the most positive experience possible for the cats.  
In practice, the project negotiated many specific ethical issues concerning the welfare and autonomy of the cats which needed to be carefully resolved.}
However, specific challenges and design details are not our primary focus here. Rather we are concerned with the eighteen-month long ethical review journey that involved a dialogue with three separate IRBs to shape the project.  

In many ways, Cat Royale was a surprising project, having emerged as the artists’ creative response to the challenge of trustworthy autonomous systems. The ethical review process was a constructive and ultimately successful one, during which many colleagues made extensive efforts to help us identify and resolve ethical issues, expending extra efforts beyond the formalities of reviewing official paperwork. Our aim is to surface this important and often hidden work of ethical review in a way that can be useful to others. Colleagues on those IRBs were shown versions of this paper as it evolved and one of the review boards invited us back to present the outcomes of the project afterwards. 

While Cat Royale was a highly unusual project, we suggest that our reflections raise wider considerations for ethical review process in HCI, especially for projects adopting multispecies \hl{and ‘beyond human-centred’}~\cite{wakkary2021things} perspectives, including those focused on animals’ direct interactions with technologies, but potentially many others that might consider the inadvertent impacts of digital technologies on animals in the wider environment. In this spirit, we contribute a new framework, \textit{beneficiary-epistemology} space, to enable future projects to position themselves in relation to their IRBs, according to whom might be considered as benefitting from the research (humans and/or animals) and to underlying epistemologies that determine what might be viewed as valid research, and so anticipate tensions likely to emerge during ethical review.  
\section{Related Work}
Below we highlight relevant previous work, which  includes general discussions of ethical review processes in HCI, and specific accounts of the challenges of approving artist-led and ACI research. 

\subsection{Ethical processes within HCI}
Engaging in formal ethical review is a routine matter for many HCI researchers, who need to justify their proposed work in terms of its benefits to society, while providing reassurance that risks of potential harm have been considered and duly mitigated. Many of us follow local or national ethical review processes as stipulated by our institutions and/or funding bodies. Indeed, publication at CHI (or any other ACM venue) requires adherence to ACM’s Publications Policy on Research Involving Human Participants and Subjects which states: \textit{``ACM authors must ensure that their human research planning, conduct, and reporting are consistent with their local governing laws and regulations''}~\cite{ACMPubBoard}. 

Aspects of research proposals that are typically of concern for IRBs include whether proposed activities might pose any direct or indirect risks for human participants (e.g. to their body, psyche, resources, reputation, privacy, relations), on what grounds said risks are worth taking (i.e. whether they are warranted by envisaged benefits) and how they will be mitigated~\cite{UKRI:2023a}. As a risk factor, plans for handling collected data are also of concern for IRBs (e.g., secure storage, anonymisation, confidentiality, destruction after a set period), as envisaged by international regulations (e.g.,~\cite{GDPR:2016}). Moreover, how participants will be enabled to provide informed consent to their involvement and to effectively withdraw from research activities is an essential consideration for IRBs (e.g.,~\cite{UKRI:2023b}). The bar is even higher when it comes to so called non-competent participants (e.g. young children, persons with limited mental capacity), who are deemed unable to provide informed consent to their involvement and who may be involved with the consent of proxies acting in the participants’ best interests (e.g. parents, legal guardians)~\cite{UKRI:2023c}.

There has also been reflection on the adequacy of current forms of ethical review from within HCI. \citet{Brown:2016:FiveProvo} offered five provocations to kick-start discussion: (i) \textit{``written informed consent does little to protect participants''}; (ii) \textit{``interventions with vulnerable populations must result in greater benefit for them than for the researchers''}; (iii) \textit{``anonymisation should be an option presented alongside co-creation of research with participants, not a default''}; (iv) \textit{``institutional IRBs delay and damage research out of proportion to any harm they prevent''} and (v) \textit{``publication of research performed with, or within, a commercial entity should be blocked until the complete dataset is made available to others''}. They propose three practical responses: differentiate between practice and law, support low-risk ethical experimentation, and think more deeply about the definition of harm. When considering the role of IRBs, they highlight how the dynamic nature of HCI research often requires changes to protocols between design and implementation; suggest that review by other disciplines can introduce \textit{``nonsensical requirements''}; and argue for \textit{``researcher and institutional responsibility rather than legalistic bureaucracy''}.  

\citet{Munteanu:2015:SitEthic} reflect across four cases of seeking formal ethical approval for HCI research with vulnerable populations including people with low literacy, parents of sick infants, soldiers, and visually impaired individuals. In contrast to Brown et al.'s provocation~\cite{Brown:2016:FiveProvo}, they highlight how dialogue between researchers and IRBs led to a satisfactory resolution and call for researchers to see boards as research partners and to engage with other disciplines when they enter new interdisciplinary territory. They propose five principles to enable the development of a \textit{``situational ethics''} framework for HCI: looking for ethical triggers, adjusting protocols in the field, dialoguing with IRBs, assembling multi-disciplinary teams, ensuring IRBs have multi-disciplinary expertise, and becoming involved in revising ethical guidelines. 
Our paper is a response to both Brown et al.'s~\cite{Brown:2016:FiveProvo} provocations and Munteanu et al.'s~\cite{Munteanu:2015:SitEthic} reflections. The multidisciplinary and multispecies aspects of our project also required us to engage with two specific areas of ethical process, discussed below. 

\hl{Broadening our perspective beyond HCI, researchers from the Social Sciences have voiced concerns about ethical review processes and the role of IRBs, especially the dangers of importing pre-emptive ethical regulation from biomedical science}~\cite{Dingwall:2008:Ethics}. \hl{One concern is that IRBs may undermine academic freedom in the interests of managing reputational risk to institutions by seeking to avoid embarrassment, litigation, and threats to funding} ~\cite{lincoln2004qualitative}; \hl{weeding out politically sensitive studies} ~\cite{moss2007if}; \hl{and employing bureaucratic process as a \textit{“serendipitous device to frustrate and deter what is considered a potential threat to an institution’s reputation or access to revenue sources”}} ~\cite{librett2010apples}. \hl{A second concern is that underlying epistemological differences may lead IRB’s to seek to make research proposals appear more conventional from their perspectives. This may arise from a lack of appreciation of alternative research paradigms and the diversity of data involved, or even prejudice regarding what constitutes valid research, and has been noted as a particular challenge for those employing qualitative, participatory, action and critical theory methods} ~\cite{lincoln2004qualitative}.
\hl{While many have voiced concerns about the role of IRBs, Stark notes a tendency for researchers to report \textit{“horror stories”}, and calls for greater mutual understanding and dialogue that respects and negotiates mutual subjectivities, including meetings in addition to formal paperwork} ~\cite{stark2007victims}. \hl{However, Martin and Inwood observe that direct discussions with IRBs can be fraught with power differences, especially when students, early career and/or untenured researchers are involved, and so call for “open discussion of the power dynamics, subjectivities, and challenges of formal ethical research structures”} ~\cite {martin2012subjectivity}.

\subsection{Ethical review process in artist-led HCI}
HCI’s turn to the cultural has brought the creativity of artists to bear on its research, for example, through activist art for sustainability~\cite{DiSalvo:2009}, improvisation techniques for disruptive innovation~\cite{Andersen:2018}, and employing discomfort for enlightenment, entertainment and sociality~\cite{Benford:2013:PerfLed}, while simultaneously extending HCI’s methods to better meet the needs of artists~\cite{edmonds2014human}. This has also challenged HCI’s ethical processes. For example,~\citet{reilly2014blending} reflect on the process of negotiating ethical approval for the public art installation Tweetris, including: not having a priori research questions that could be answered through controlled experimentation; difficulty modifying approved protocols as the artwork evolved; and obtaining informed consent for walk-up participation. 

A gathering of artists, HCI researchers, and ethicists to share experiences of ethical challenges and processes highlighted distinctive artistic perspectives contrasting with those imported from Medical Science via experimental Psychology~\cite{Benford:2015:Ethical}. Participants reported how tensions arose from multiple-overlapping ethical frames: artists typically operate according to the ethical processes of the professional artworld, but these are typically quite different from those of institutional research. The relationships between these frames can become particularly confused when the artists choose to ambiguously blur the relationship between art and science, for example, adopting the persona of a scientist in a performance. A second tension concerned the idea of doing ethics throughout a project, including ``on the way out’’, referring to the engagement of audiences in ethical debate after the experience, and possibly with a lighter touch beforehand. They further noted how artists draw on their carefully honed professional judgement to dynamically negotiate the boundaries of consent and withdrawal, which are typically personal, fluid and contingent, \hl{while recognising that this can be challenging and difficult territory}. Finally, participants highlighted the challenge of data veracity especially in situations where artists appear to introduce elements of science, a finding that was reinforced by a subsequent study of the creative use of physiological ‘biodata’ in promotional filmmaking~\cite{Reeves:2015:Biodata}. 

\subsection{Ethical review and disciplinary assumptions in ACI}
\hl{HCI's growing interest in Animal-Computer Interaction and turn to designing for more than human-centred worlds} ~\cite{wakkary2021things}, \hl{raises distinctive ethical challenges concerning the involvement of animals and potentially other non-human stakeholders}. Research involving animals is usually the responsibility of institutional ethical review bodies and is informed by relevant legislation (e.g.~\cite{EC:2010,ASPA}). An underlying assumption of such legislation is that being involved in research is rarely in the animals’ interests, on the grounds that they do not have the intellectual capacity to consent to procedures that may harm them, and that the individuals involved are effectively used for a greater good, which may include benefits to the environment, to species of flora and fauna and, above all, to humans. Consistent with this assumption, the principles of Replacement, Reduction and Refinement (3Rs)~\cite{russell1959principles}, whose application is now generally regarded as the gold standard of humane research, require that: whenever possible, animals be replaced by other methods or simpler species be used instead of more complex ones; the number of animals involved be reduced to a minimum necessary for statistical power; and protocols be refined to minimise suffering before, during and after procedures. Although the protection of research animals is ultimately subordinated to the integrity of scientific procedures~\cite{Mancini2022relevance}, researchers must make a compelling case for their use by highlighting the potential societal benefits of planned work and, critically, by demonstrating the quality of their scientific method. This means formulating clear hypotheses that can be validated by falsification of reasonable alternatives and accumulation of empirical evidence, with emphasis on quantification and reproducibility, to formulate theories that have predictive power (e.g.,~\cite{popper2005logic}). In many contemporary societies, the scientific method (with the scientific evidence it produces) constitutes, at least in principle, the foundation for policy development, on the grounds of its presumed objectivity. Ultimately, because objectively measuring animals’ intellectual capacities and emotional experience is hard, in legislation there remains an assumption that, while they can be used as objects of study, animals cannot be regarded as subjects capable of understanding, consenting to and participating in research activities. 

In marked contrast, and paralleling human-centred values inherited from HCI, ACI takes a fundamentally different approach to the ethics of research involving animals. At least in principle, ACI adopts an animal-centred perspective grounded not only in animal welfare frameworks (e.g.,~\cite{Vaataja:2014:AnimalWelfare,webber2022welfare}) but also in political philosophies of multispecies justice, with an emphasis on animals’ agency, capabilities and dignity (e.g.,~\cite{Mancini:2023:Politic,Linden:2023:AnimalCentered}). Consistent with this perspective, ACI researchers have proposed ethical frameworks for the involvement of animals in research(\cite{Vaataja:2013:EthicalIssues,Hirskyj:2016:Ethics,Mancini:2017:ACI}) as well as approaches to support researchers’ ethical engagement with individuals during research activities~\cite{ruge2022ethics} . Akin to ethical principles for biomedical research with vulnerable humans~\cite{beauchamp2001principles},~\citet{Mancini2022relevance} have proposed the principles of Relevance, Impartiality, Welfare and Consent as a counterpoint to the 3Rs, aiming to problematize the 3Rs’ applicability to ACI research and to complement the 3Rs’ application to animal research beyond ACI. These principles require the research to benefit partaking animals, to ensure the highest standards of treatment regardless of species, the protection of partakers’ wellbeing always, and the provision of opportunities for them to consent or dissent to their involvement (in the complementary forms of \textit{mediated} consent provided by legal guardians and \textit{contingent} consent provided by the animals themselves). These protections are not predicated on pre-established animal capacities, but rather on animals’ role as research participants and autonomous agents regardless of any capacities of sentience, abstraction, rationality or language.  

The recognition of animals as autonomous agents and legitimate participants with a stake in research is aligned with critiques of anthropocentrism (e.g.,~\cite{Moore:1906,kohn2013forests}), which decentralize the human by defining agency, not in terms of intellectual capacity, but as the capacity of interconnected organisms (and even non-organic actors~\cite{latour2007reassembling}) to influence and change one another~\cite{Haraway:2008:Species}. Furthermore, these critiques reject the idea of scientific objectivity and question the reliability of scientific knowledge, instead highlighting the situatedness of knowledges, which are constructed as individuals engage with the world and attribute meaning to worldly interactions~\cite{Haraway:2016:Situated}. In line with these positions, alongside work that applies the scientific method (e.g.,~\cite{Zeagler:2014:GoingDogs,Byrne:2017:Dogs,Majikes:2017:Balancing,Paci:2017:RoleEtho}), ACI researchers have drawn on fields such as anthropology (e.g.~\cite{Sadetzki:2022:Leashing}, anthrozoology~\cite{Westerlaken:2016:Becoming}, multispecies ethnography (e.g.,~\cite{Mancini:2012:Interspecies}) and ethnomethodology (e.g.,~\cite{Aspling:2017}), or speculative design~\cite{Lawson:2015:Prob}, to frame multispecies notions of participation and co-design~\cite{Mancini:2018:Emerging}, and to propose participatory methods for dealing with interspecies differences and communication barriers~\cite{Robinson:2014:Canine} in ACI research. 

\hl{However, critiques of ethics of relationality and entanglement}
~\cite{giraud2019comes} \hl{highlight how situatedness inevitably implies exclusion, whereby ethical and methodological choices that define research projects exclude other possibilities while being necessary to enable researchers to address specific questions. This also concerns decisions on whether and how to involve animals in research, which are made on the animals' behalf and inevitably limit the animals' agency. In this regard, ACI researchers} ~\cite{mancini2018emerging} \hl{have highlighted how, in multispecies interaction design, research set-ups that foster \textit{semiotic}, \textit{volitional} and \textit{choiceful} engagement can support a process of progressive orientation towards animal-centred outcomes that may never be fully reached, but that can be incrementally approximated, despite inevitable limitations. At the same time, ACI researchers} ~\cite{mancini2022politicising} \hl{have stressed the importance of considering the political implications of ACI research beyond specific projects. This requires being mindful of how research taking place within limitations imposed by socio-economic contexts that are not animal-centred may impact animals beyond specific ACI projects. Articulating and negotiating the ethical boundaries between the project and these wider contexts was a key challenge for Cat Royale.}  
\section{A Brief Introduction to Cat Royale}
Cat Royale was part of the UKRI Trustworthy Autonomous Systems Hub, a national research programme in autonomous systems which included a Creative Programme for artists to make new work to engage the public and media. Blast Theory were the programme’s Creative Ambassadors, charged with delivering a flagship artwork, and Cat Royale was their response. From the outset, the project was driven by two goals. The first was to surface the matter of trust in autonomous systems by exploring whether people would trust a robot to look after their loved ones as represented by cats. The second was to consider how autonomous systems might benefit companion animals by exploring whether a robot could successfully play with the cats.  

The artists deliberately \hl{and ambiguously} framed Cat Royale as intending to provide a ‘utopia’ for cats. On the one hand their aim was to create a bespoke and luxurious environment that would cater to the cats' various needs~\cite{Ellis:2013:AAFP}, including using a robot to enrich their lives through play, and using AI to learn their preferences and measure their resulting ‘happiness’. On the other, this framing was deliberately intended to provoke questions in the viewer: what would a utopia for cats be like? How could AI measure their happiness? Could we trust the robot and the AI with the cats? And how might this translate to humans? We now briefly summarise the design and experience of Cat Royale as background to help better understand its journey through the ethical approval process. Further overviews of Cat Royale can be found in~\cite{Schneiders:CatRoyaleTAS:2023,Schneiders:2024:Interactions,Schneiders:2024:altHRI} while~\cite{Schneiders:2024:CatRoyaleRobotWorlds} provides a detailed account of the design of the robot and the enclosure, concluding that it is important to carefully design the world in which a robot operates.

\subsection{The design of Cat Royale}
The design of Cat Royale emerged iteratively. At the core of its final version was a purpose-built enclosure designed to be inhabited by a small family of three cats (a parent and two offspring). The cats lived here for six hours a day, split into two three-hour long periods with a break of several hours between. It provided ample space, with dens, perches, walkways for jumping, sleeping and observing; as well as a scratching post, a water fountain, cat grass, feeding stations and litter trays to be suitable for the three cats, as advised by experts in cat welfare. Its design was specifically informed by the sensory, cognitive and physical characteristics of cats, for example, in the choice of colours suitable for dichromatic vision, surface coverage suitable for clawed limbs, etc. The enclosure was also designed to appear luxurious to human observers, through its visually eye-catching bespoke interior design.  

Inside the enclosure was a robot arm, securely mounted to the floor adjacent to an accessory rack with four slots, on which various attachments holding toys could be placed, ready for it to pick them up and deploy them. Example activities included dangling and waving various soft toys on strings, slowly dragging a feather boa across the floor and dropping balls into a tube from which they would emerge to roll across the ondulated floor. The robot could also offer treats (kibble and meat sticks) on a tray. The arm was a Kinova Gen3 lite, a lightweight robot arm equipped with a non-interchangeable two-finger gripper. This was chosen to be strong enough to be able to lift the toys off the racks and wield them, but weak enough that it would not harm the cats should it accidentally collide with them. Custom toy attachments enabled the robot to retrieve the toys from the rack when required, and store them back out of the cats’ reach when no longer needed. The attachments adhered magnetically to the rack to minimise the risk of the toys being accidentally dropped on or deliberately ripped down by the cats. 

\begin{figure*}
    \centering
    \includegraphics[width=\linewidth]{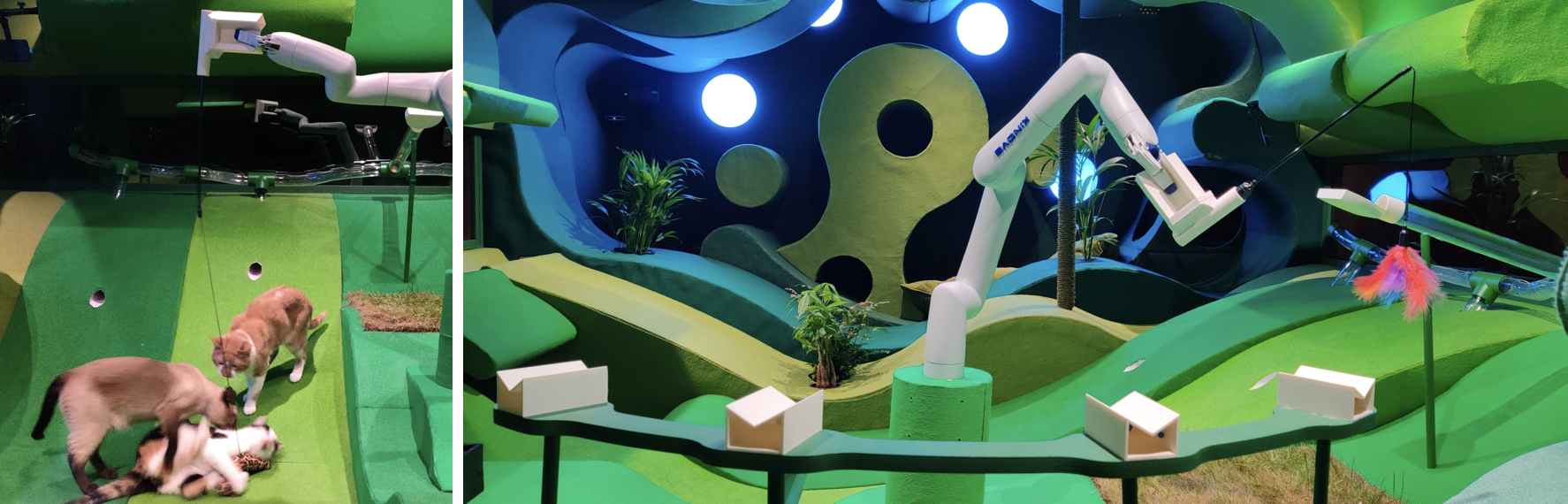}
    \caption{The the three cats (Clover, Pumpkin, and Ghostbuster) playing with a toy (left) and the robot arm lifting the feather helicopter toy from the rack (right)}
    \label{fig:ThreeCats}
    \Description{On the left of this figure we see an image of three cats playing together with a small soft toy that is being dangled on string by a robot arm that hovers above them. On the right we see a picture of the enclosure within which this happened showing its striking decorative design and with the robot arm positioned at the centre, The robot has just lifted a toy attached to a rod from off of one of four raised racks in front of it.}
\end{figure*}

Every ten minutes, the arm would attempt to engage the cats by picking up a new toy or offering a treat. This would involve executing a programmed movement sequence for the object that the artists had designed and tested beforehand. This would be combined with various generically useful actions such as picking attachments off the racks, replacing them, and the robot folding itself down into a safe ‘off’ position when inactive. 

Outside the enclosure, the artists monitored the cats from behind one way glass mirrors, manually rating their engagement with each game using the Participation in Play scale~\cite{Ellis:2022:Pip}, a tool used in Veterinary Science to measure behavioural indicators of wellbeing in cats through ordinal scales that rate aspects of play, such as playing enthusiastically or tentatively, playing while moving or stationary, watching toys passively without engaging, appearing disinterested, or retreating. The results were fed into an AI ‘decision engine’ that attempted to learn which toys and treats the cats preferred, and that would then recommend the next toy or treat to be offered, which the artists could accept or reject (recording the reasons for their decision).  

A cat Welfare Officer, with post-graduate level qualification in animal behaviour and working professionally in animal welfare, continually monitored the cats for any signs of stress while they were in the enclosure and in the studio, and trained other members of the team to do so too. They took notes on the cats’ behaviour and regularly completed the Cat Stress Score~\cite{kessler:turner:1997}, feeding their observations back during a daily meeting of project management team that included artists, veterinarians, and selected researchers, who continuously reviewed how the project should be adjusted. Strict protocols were drawn up for closing several layers of doors so that the cats could not wander into the unfamiliar and potentially dangerous environment beyond the enclosure. 

A human robot-operator oversaw the operation of the robot, triggering its pre-programmed sequences and sometimes having to improvise movements to untangle it from unexpected situations. They had to continuously hold down a button (a ‘deadman’s switch’) to enable the robot to move, preventing it from performing any movement without explicit authorisation from the operator, and controlled an additional emergency stop button, situated directly in front of them, that would instantly shut down the power to the robot in an emergency. These various measures implemented a strict supervisory control paradigm~\cite{sheridan1992telerobotics} whereby at least one human always observed the operation and was responsible for ensuring safety. A toy-wrangler was tasked with loading toys onto the rack and maintaining them and the enclosure. 

Finally, the enclosure was constantly filmed using eight embedded cameras, whose outputs were mixed by an experienced television vision-mixer to produce video material with annotations labelling the games, cats and estimated happiness scores. This was subsequently edited into an seven-hour long movie to be shown to the public as a touring installation for art galleries as well as a series of short daily highlights released on YouTube. 

\subsection{The experience of Cat Royale}
Over the course of twelve days the robot experimented with over 500 games and treats. The installation run safely and the cats did not show signs of stress; at no point were they withdrawn or was the emergency stop button deployed. The cats quickly settled into the environment and usually engaged enthusiastically with both toys and treats. They mostly avoided direct physical contact with the robot arm itself, focusing their attention on the toys. A typical play sequence was for one or more cats to observe from a high perch as the robot retrieved a toy, before descending to crouch on the floor, before then physically engaging with the toy (biting, pawing or batting) for up to several minutes. Over time the cats became more confident and assertive with the toys. Clover, in particular, was persistent in her interactions with the robot, eventually working out how to exert sufficient force that the robot-operator had to trigger the robot to let go of the toy, allowing her to drag her prize away across the enclosure and out of its reach. The robot often got into various tangles as strings got wrapped around it or balls dropped in the wrong places, which required manual improvisation from the operator to resolve. The system learned that the cats preferred certain toys; the ‘feather helicopter’ was a favourite. Generally, all the cats made ample use of the space and of the resources within it, for example, lounging in the elevated dens, patrolling the suspended walkways and rolling surfaces, availing themselves of the concealed litter trays, and savouring the fountain's running water. 

The edited film has been exhibited three times so far, there has been national press coverage, and discussion on social media, with the artists fielding many questions, but nothing to suggest an outcry or backlash against the project.  

However, the point of this paper is not to break down the detailed design or experience of Cat Royale, nor to examine in what respects it did, or did not, succeed for the cats or wider audience. Rather, we reflect on how the project negotiated a complex ethical process and various ethical challenges during its design and development.
\section{Cat Royale's Ethical Review Process}
The ethical design and review process passed through four distinct stages: initial discussions of ethical challenges, processes, and available expertise within the project, followed by engagement with three separate IRBs: the University-wide \textit{Animal Welfare and Ethical Review Body (AWERB)}, the \textit{Computer Science Ethics Review Committee (CSREC)}; and the Vet School’s \textit{Committee for Animals and Research Ethics (CARE)}.
\hl{We now explain how each of the stages unfolded, focusing on how they revealed different perspectives on both the beneficiaries of the project, and the underlying validity of its methods and resulting knowledge.}

\subsection{Initial internal ethical discussions}
\hl{It was understood from the outset that Cat Royale was a complex ethical proposition. The artists’ clear intention in proposing the idea was to create a situation in which people would be invited to explore their feelings about trust and autonomy in relation to AI and robots. Getting cats to live and play with a robot in a so-called ‘utopia’ was a carefully pitched provocation in this regard, and one that many people found intriguing but also somewhat uncomfortable.   

We recognised the importance of involving expertise in cat welfare and behaviour from early on and so invited experts in animal-computer interaction and ethics, and veterinary behavioural medicine and feline behaviour to the team. We also solicited advice from our national society for animal protection and established a fifteen-strong Audience Advisory Panel (AAP) with varied interests in the arts, technology and cats, who met five times as a group.

Early discussions among this expanded team surfaced four key ethical challenges. The first concerned the choice to work with cats. This was in part an aesthetic choice as cats were felt to combine a compelling cuteness with an enigmatic sense of autonomy and  aloofness alongside an established Internet presence. However, it was also a practical choice due to their moderate size.}

Our second challenge was whether the work (and cats) should be located at a public gallery or whether the team and robot should travel to the cats. Ultimately, we decided to locate the enclosure, the cats and their owner at the artists’ private studio, while public venues hosted video installations.

The third challenge concerned the idea that the system would somehow measure ‘cat happiness’. The artists felt this to be an ambiguously provocative proposition for the audience; the computer scientists developing the technology were looking for something that could be readily implemented; but animal researchers worried about the scientific validity of an overarching and simplistic idea of happiness. Ultimately the decision was taken to adopt the Participation in Play scale ~\cite{Ellis:2022:Pip} with the rationale that it would best represent the AI’s own view of happiness – i.e., that its task was to play with cats to make them happy.

Our fourth challenge concerned the autonomy of the cats. To what extent would their engagement be voluntary and consensual? Could they choose to withdraw from the robot or the enclosure? Ultimately, the enclosure was designed to afford them ample opportunities to autonomously approach or avoid the immediate vicinity of the robot, while the decision to remove them from the enclosure rested with the Cat Welfare Officer, their owner, and the artists. \hl{The challenge of constraining the cats' autonomy was to resurface throughout our subsequent discussions with IRBs, and to some extent, ultimately remained unresolved as an inherent tension in the work intended to provoke the audience to consider the wider implications of AI.}

\subsection{Discussions with the Animal Welfare and Ethical Review Body (AWERB)}
In our country there are strict laws and regulations governing the use of animals in research and the work of all establishments that conduct research with animals is overseen by AWERBs to ensure that projects comply with the law and that the animals involved are duly protected. Our AWERB includes veterinarians, animal welfare officers, scientists and lay people, and advises staff on the acquisition, welfare, housing and use of animals in research.  Their remit includes considering \textit{``the scientific benefit of the work weighed against the cost to the animals involved''} and \textit{``the application of the 3R's (Reduction, Refinement and Replacement)''}. 

AWERB members were generous in engaging in informal discussion to help us shape our formal ethics application. \hl{One issue that arose in relation to the wider framing of the project concerned its intended beneficiaries}. As one advisor informally observed: \textit{``...you state the primary goal of the work is cat welfare, but that doesn’t come through very clearly in the background where the purpose appears to be about exploring the public’s ambivalence around automated systems''}. They further noted that \textit{“at the moment the benefits to the cats don’t come through ‘til later in the case, and it could be read as being primarily an art installation.”} 

\hl{A second question concerned the validity of our methods and the knowledge we would produce.} Our initial formal application to AWERB received the feedback that the application could not be approved because: \textit{``there was no clear experimental design''}, and \textit{``the type of data that you want to collect needs clarifying''}. 

In our revision and resubmission, we rewrote our objectives, trying to clarify and balance potential benefits to cats and humans. We also tried to clarify our epistemological and methodological position as a response to concerns about ‘experimental design’ and ‘data capture’, explaining that our project was intended to be a combination of (i) a technology development project that \textit{``follows a co-design process that involves stakeholders (in this case humans and animals) early on and unfolds through a series of iterations''} and (ii)  an artwork seeking to \textit{``engage a wider public in reflecting on the consequences of introducing new autonomous technologies into the home; on the welfare of and their relationships to their companion animals; and to promote the idea that such technologies need to be designed for and with animals.''} We further explained that this was not intended to be a \textit{``scientific project that involves a series of cat research experiments, or that employs autonomous systems to test hypotheses about feline psychology and behaviour''}.

AWERB’s response to our revised submission was to allow the project to proceed, at least to the next stage of being able to engage cats with the robot as part of the unfolding design process, while requesting updates from us alongside pro-active feedback to help the University’s media department respond to any public or media queries. They \textit{``appreciated that the data collected may be largely subjective, not objective''} but noted that \textit{``there was still uncertainty and concern over the ‘art versus science’ divide and the possibility of reputational damage for the university''}. \hl{The latter comment led us to clarify our communications strategy by nominating the artists and principal research investigator as media and public contact points, while other researchers maintained their anonymity (at least during the initial filming and public engagement phase); anticipating public and media questions and planning responses; undertaking media engagement training; and briefing the University press office so that it could handle inquiries.}

\subsection{Discussions with the Computer Science Research Ethics Committee (CSREC)}
The next step in our ethical review process was to seek approval from CSREC for the more human-facing aspects of the project. We completed the CSREC paperwork, including the AWERB paperwork as an appendix alongside a Data Management Plan, draft Cat Welfare Protocol, and \hl{our communications strategy}. CSREC’s initial feedback requested further detail about the design of interviews and surveys, and procedures to protect people’s identities in social media posts; and asked for clarification about the ongoing involvement of cat welfare experts (including their presence onsite) and various details around the involvement of the cats and operation of the robot. Their feedback also highlighted potential reputational risks to the University and the wider Trustworthy Autonomous Systems research community, especially around potential negative public and media reaction to the AI aspects of the project, with an explicit request that \textit{``the study team reflect on the potential relationship between negative responses to the project and the project’s own framing''}. They observed that: \textit{``It is possible that any media/protest group attention might expand outwards from the welfare of the individual cats into discussions around AI being forced onto animals without their consent (and humans are next)''}, further observing that \textit{``... media presentations around AI often fall into extremes of either utopia or dystopia, which prevent much nuanced discussion occurring and can cause discussion to be polarised''}. 

\hl{These comments revealed the need for clearer communication of the artistic framing of the project, especially with regard to the balance between benefiting cats versus provoking questions in humans. In response, we} wrote a document explaining how the artistic framing of a ‘cat utopia’ involved a deliberate use of ambiguity to provoke people to reflect on wider implications of AI, and how this built on a history, within both art and HCI, of harnessing ambiguity (including the deliberate use of over-precision (citing~\cite{Gaver:2003:Ambiguity}). However, we also recognised that there should be no ambiguity about our position with regards to the cats: that we should better communicate how we were focused on trying to provide them with the most positive enriching play experience we could, while meeting their welfare needs.  

CSREC was sufficiently satisfied with our response to be able to grant approval for the human-facing aspects of the project. However, it expressed concern that there were still unresolved issues around the treatment of the cats and the potential for wider risk to the University beyond its remit. At this point discussions progressed to a Science Faculty level where it was decided to seek a further round of approval from an additional committee. 

\subsection{Discussions with the Committee for Animals and Research Ethics (CARE)}
Our engagement with the CARE in the Veterinary School occurred in the final month of project development and so presented an opportunity to review our design in its final form after several iterations of testing and development had occurred. Our submission to CARE included a Cat Welfare Protocol that set out procedures to be followed to ensure the welfare of the cats. Much of CARE’s feedback focused on this, with requests to clarify aspects such as providing a suitably long habituation period, whether the cats would also be monitored when outside of the enclosure and in their living quarters within the art studio, whether they were already part of a social group, assessment scales to be used, and the level of experience or training of the Cat Welfare Officer. These issues were relatively easy to accommodate in the final protocol. 

A more challenging issue (that had been present since our early internal discussions) concerned the cats’ autonomy, especially their ability to withdraw from the enclosure. It was clear to all that the cats should be promptly removed from the enclosure if showing any signs of fear or stress. Less clear was what should happen if they appeared to be bored by the robot/enclosure and more interested in what was happening outside. Should they be able to leave the enclosure at will via a cat door? Given safety concerns about the environment immediately beyond the enclosure (narrow, dark, full of cables and computers, and with a door to the wider world that might be left open for ventilation), but also a desire to keep the artwork running whenever reasonable, it was decided, after considerable discussion, that the door to the enclosure would be closed when the cats were inside, that this would only be for 3 hour stretches, and that the Cat Welfare Officer and/or owner would be able to instigate removal of the cats if deemed important for their welfare.  

\hl{Like AWERB before, CARE was also concerned about the validy of our methods and the generality of the knowledge they would produce; specifically that the project might be seen by the public and by other scientists as an attempt to undertake scientific experiments into animal behaviour}. The committee requested that the project should be portrayed as a \textit{``one-off exercise and that the behaviour recorded cannot be generalised to other cats, single cats, other breeds, other social groups''}. We subsequently coordinated with CARE to agree the final wording on the project website and in press releases. 
\section{Discussion: Revealing Ethical Tensions}
We identify three broad reasons why Cat Royale needed to navigate such a complex ethical review process. First, it was a complex project involving many moving parts — animal, human and robot — and so needed to draw on varied expertise to surface and resolve diverse ethical challenges. Second, its iterative design-led approach required a similarly iterative ethical process that moved from surfacing broad concerns to ultimately shaping and approving fine details (reflecting Munteanu et al.’s~\cite{Munteanu:2015:SitEthic} and Brown et al.’s~\cite{Brown:2016:FiveProvo} previous observations concerning the dynamic nature of HCI research in relation to ethical review). Third, the ethical review process had to negotiate underlying differences between the diverse disciplines that became involved. These differences were foundational to how different parties approached the process. They were also largely hidden, arising from assumptions about how research should be done and/or how others might view research as described. This leads us to propose a framework to help make explicit the disciplinary landscape within which complex multi-disciplinary and multispecies research projects must negotiate ethics. 

\subsection{Beneficiary-epistemology space}
We define this landscape through two orthogonal dimensions that, on reflection, we realise underpinned the many ethical discussions we were having.  

\textbf{Beneficiaries} – considers who benefits from the research and correspondingly who does not, including who assumes the risks of potential harms; in other words, anyone who could be regarded as a stakeholder. We divide beneficiaries into human and animal. \hl{Human stakeholders may span research participants (e.g., the cats' owner), target end users (e.g., the public audience), but also many others with a stake in the project (e.g., the artists, University researchers and their academic colleagues).} Similarly, animal stakeholders may include those directly involved in research activities, those targeted by the research, and others in the wider ecosystem. The most challenging ethical tensions typically arise when those who might potentially be harmed are not the same as those who might benefit, especially if they are unable to consent or effectively dissent.   

\textbf{Epistemology} – considers what different disciplines regard as valid knowledge and methods for attaining it. With regards to multidisciplinary research, Moon et al.~\cite{Moon:2021:Five} discuss how epistemologies may be broadly situated between two opposing poles: on the one hand,positivism/objectivism emphasises objective knowledge gained by observation and validated by experimentation; on the other hand, interpretivism/subjectivism  emphasises that the beliefs of individuals or communities constitute legitimate knowledges, and that researchers are inherently part of the research and so can never be objective. As the authors highlight, tensions are likely to arise when research involves multidisciplinary contributors who subscribe to different epistemologies and, thus, different beliefs as to what constitutes knowledge and how knowledge is developed. 

Against these two dimensions important tensions pertaining to ethical review become apparent. Figure two shows the \textit{beneficiary-epistemology} space of Cat Royale, populated (to a first approximation) with the project (yellow box), our three ethical IRBs (green), and important tendencies within the underlying disciplines that they serve (blue labels). 

\begin{figure*}
    \centering
    \includegraphics[width=.7\linewidth]{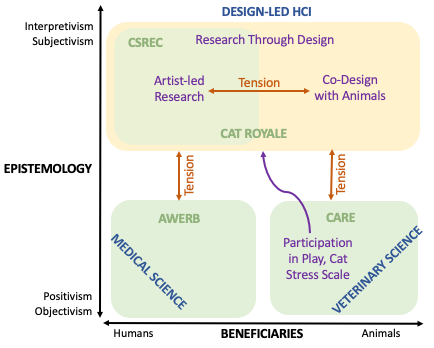}
    \caption{Beneficiary-epistemology space reveals the disciplinary alignment of Cat Royale and its three IRBs and hence the tensions in its ethical review process.}
    \label{fig:BenEpSpace}
    \Description{A diagram in the form of a two dimensional taxonomy. The horizontal dimension is labelled beneficiaries, and runs from humans on the left to animals on the right. The vertical dimension is labelled epistemology, and runs from positivism and objectivism at the bottom to interpretivism and subjectivism at the top. Across the top half of the diagram is a shaded area representing the Cat Royale project. At the top left and so overlapping with this, is a smaller area labelled as the Computer Science Research Ethics Committee. At the bottom left, is an area labelled Animal Welfare and Ethical Review Body. At the bottom right, is an area labelled Committee for Animal Research Ethics. There is a label bottom left saying medical Science. There is a label bottom right saying Veterinary Science. There is a label middle top saying Design-led Human-Computer Interaction. A line on the diagram show an internal tension within the project and two further lines show external tensions between the project and external review boards. There is a labelled arrow to show how the Participation in Play Scale and Cat Stress Scale method were imported from Veterinary Science into the project.}
\end{figure*}

We position Cat Royale as spanning the top-half of the space. The project intended to directly benefit both humans (the public audience) and non-humans (the cats), albeit in quite different ways; the former through encouragement to reflect on AI and the latter through a positive play experience. It also involved other human stakeholders who stood to benefit and/or carried risk, including the artists, cats’ owner, robot operators and wranglers, and the wider research team. Adopting a very broad view of the surrounding ecosystem, animal stakeholders included the three cats, but potentially also their siblings (who did not participate and were temporarily deprived of their company) and even the artists’ dog (who was banned from the studio space for the duration). Turning to epistemology, Cat Royale followed an interpretivist and subjectivist design-led approach consistent with Research Through Design. However, as we discuss further below, it actually combined two distinct methods that were themselves in tension, artist-led research and co-design with animals. And while it did not claim to ‘do science’, it clearly imported methods from Veterinary Science in the form of the Participation in Play Scale and the Cat Stress Scale which it applied for its own purposes. 

We also position our three IRBs within this space. CSREC routinely deals with research involving human participants, which is expected to benefit humans. They are experienced in assessing design-led HCI research, including artist-led and co-design projects, but did not have experience of projects involving animals. CARE deals specifically with veterinary research, which typically aims to benefit animals and employs scientific methods. AWERB has a broad remit concerned with the welfare of any animals used for any research across the University and implements a legally mandated regulatory function. Although this applies to any research involving animals, it is rooted in the use and protection of animals predominantly in medical research, which typically applies scientific methods for human benefit and in which animals are instruments in the scientific apparatus. Both CARE and AWERB primarily consider the integrity of scientific processes when judging the balance of benefits and risks to either humans or animals, and compliance with legislation. They are less familiar with the design-led, reflective, and human- and animal-centred research approaches common in HCI and ACI.   

Finally, we also label some broader underlying disciplinary trends on the landscape. Veterinary Science leans towards research that often benefits animals (though it does not necessarily benefit research subjects and may ultimately benefit humans, if we consider animal farming or owning pets) and that critically employs scientific methods. Medical Science leans towards research that benefits humans (often at the cost of harming animals) and that also typically employs scientific methods. HCI traditionally seeks to benefit humans and is epistemologically broad, with foundations in the science of experimental psychology, but with later ‘waves’ turning towards interpretivist and subjectivist epistemology. The design-led HCI approach that we were following appears in the top half of the space. ACI aims to benefit animals (both research participants and end users) and, like HCI, draws on a range of methods, including those derived from the experimental tradition of Veterinary Science and from HCI’s design-led approaches. This is, of course, a simplified portrayal of what are complex disciplines that may themselves accommodate multiple epistemological stances and related methodological approaches. As mentioned above, HCI research can involve scientific methods such as lab-based experimentation. Medical and Veterinary research may accommodate perspectives from the social sciences, humanities and even the arts (e.g., in Health Humanities and Veterinary Humanities). However, we maintain that our observations highlight significant general orientations in these various disciplines that are useful for anticipating their ethical perspectives.  

Positioning the project, ethical IRBs and underlying disciplines within \textit{beneficiary-epistemology} space in this way, serves to reveal key tensions that had to be negotiated throughout the ethical review process. 

\subsection{Internal tensions within Cat Royale}
First, we recognise that the project negotiated significant internal tensions. Previous discussions in HCI have considered the complexities that arise from overlapping ethical frames that blur the relationship between art and research and this was also the case here. The overarching (or outer) framing of the project was to benefit humans by employing artist-led methods to provoke audiences to reflect on trust in autonomous systems. However, this was to be achieved through a distinct inner framing that involved applying animal-centred co-design to enable a robot to play with cats. These partially overlapping ethical frames raised confusion during ethical review. The idea of provoking audiences to reflect on trust in robots and AI implied that there might potentially be some risk to the cats, at least in theory. Would the artists need to create or at least imply situations that raised a degree of uncertainty about risks to the cats? Might it appear that the cats were being exploited for the purposes of art or even base entertainment, as if \hl{within} some kind of circus?  

This was exacerbated by methodological differences wthin the project. While both Performance-led Research~\cite{Benford:2013:PerfLed,Benford:2015:Ethical} and Participatory ACI~\cite{Westerlaken:2016:Becoming,Mancini:2018:Emerging,Chisik:2020:Politics} are design-oriented methods that share a common underlying subjectivist/interpretivist epistemology, there are important differences between them. Artist-led design relies on the professional judgement of artists to make appropriate ethical choices and carefully judge boundaries~\cite{Benford:2015:Ethical}. Their artworks may be transgressive, sometimes pushing personal boundaries, making their ability to judge whether participants consent to or wish to withdraw of paramount importance. Judging the cat’s boundaries in this respect was a new and difficult challenge for the artists, requiring input from animal welfare experts and the cats’ human owner. Conversely, ACI’s approach of co-designing with animals considers them to be partners in a co-design process, notwithstanding contextual limitations and exclusions. The ACI ethics frame required that their autonomy be prioritised to enable them to provide or withdraw contingent consent to their involvement and, thus inform the design of future cat-robot interaction experiences~\cite{Mancini2022relevance}. However, allowing the cats to exit the enclosure at will could have endangered them; since according to ACI research ethics threats to animals’ physical and psychological integrity must be avoided at all costs~\cite{Mancini:2017:ACI}, there was a case for prioritising the cats’ safety over their autonomy. Moreover, the cats’ exit would have effectively put an end to the aesthetic experience of the audience, thereby defeating the project’s artistic goals. At the same time, it was important that the audience could focus on the aesthetic experience and the questions this provoked, without being distracted by concerns over the situation of the individual cats. \hl{These contextual demands and the resulting constraints inevitably excluded possibilities} ~\cite{giraud2019comes} \hl{for the cats' agency (e.g. what they might have expressed had they been able to enter and exit the enclosure at will), but at the same time it was only through these limitations that the cats could be given representation and agency within the project and, thus, the possibility to influence its outcomes} ~\cite{mancini2018emerging} \hl{and the future of technology that might impact them, the species they represented and other animals. } 

While understandably challenging for ethical review, the deliberately ambiguous framing of the work and clash of methods were core to the artistic proposition that, through a sincere and comprehensive attempt to design a robot to benefit cats, the artists could simultaneously surface deeper issues about the role of AI in society. However, it did require us to write a document that carefully articulated this proposition for our IRBs.  

\subsection{The external tension with Veterinary Science}
Our project aspired to have a productive relationship with Veterinary Science. We were keen to draw on expert advice on cat welfare and behaviour, make use of tools such as the Participation in Play and Cat Stress Scale, and even hoped that our research findings might be of wider interest to their community. However, while we shared a common interest in the cats as beneficiaries, there was confusion about our approach. Why make an artwork rather than undertake a series of lab experiments? 

Underlying this were epistemological differences between our design-led approach to ACI and the desire for the replicability of research protocols and generalisability of findings from Veterinary Science. This became evident in our struggles to clarify hypotheses and data collection methods, and ultimately the requirement for us to publicly clarify that this was a one-off artwork, not a repeatable experiment. With hindsight, this is understandable. While we were aspiring to benefit cats, we did not subscribe to the widely accepted positivist epistemology or apply methods recognised by this community. Thus, even if we did not actually harm the cats, it was not clear on what objective grounds we could justifiably claim to benefit them or cats in general.  

To further confuse matters, we were evidently trying to apply techniques developed within Veterinary Science, but potentially in an unprincipled way that might be seen as bogus science, something that could be questioned by the wider scientific community. This reflects HCI’s previous discussions of the challenges that arise concerning ‘veracity’ when artists engage with science, for example, when presenting themselves as scientists within an artwork~\cite{Benford:2015:Ethical}.  

Given that this was intended to be a very public project and so would likely attract widespread scrutiny (including from other veterinary scientists), there was a risk to the University’s wider reputation for veterinary science. In the worst case, were serious incidents to occur (e.g., if the cats \hl{were to have been} harmed somehow), the resulting furore could impact the ability of the University’s veterinary science community to conduct their own research in the future. It is not difficult to imagine (though no one said so) how we might appear to be ‘tourists’, visiting another discipline, taking away some ‘souvenirs’ (methods), and potentially leaving a trail of damage in our wake, while we quickly moved on to new destinations. Indeed, this is perhaps a wider consideration for a field such as HCI that routinely turns to other disciplines for inspiration and knowledge, whose methods it may then appropriate for its own purposes. Ultimately, this tension was resolved through a dialogue with CARE that led to a clearer internal and public recognition of epistemological differences, and through the recruitment of a recognised veterinary researcher and their expertise in cat behaviour, who would be able to advise on the project. 

\subsection{The external tension with Medical Science}
At first sight Cat Royale might appear to be far removed from medical science. However, we were aware of potential confusions here that we wished to avoid. First, we recognised a potential reading of the project that we might be seeking to experiment on animals (at some risk to them) in the interest of benefiting humans (even if not medically). Second, while AWREB does not exclusively deal with the use of animals in medical research, its remit includes ensuring compliance with relevant legislation that does arise from this, especially with the principles of the 3Rs, which is regarded as insufficient by ACI ethical standards~\cite{Mancini:2017:ACI}.  Moreover, medical research is largely grounded in a positivist scientific epistemology, so that there were potential tensions around both beneficiaries and epistemology.  

In contrast to our engagement with Veterinary Science, this was not a disciplinary relationship that we sought but, rather, it was one that we needed to anticipate and avoid. We did not wish to borrow from or do medical research but needed to manage the risk that we might be viewed by the public as though we were experimenting on animals or be judged in this way during the ethical approval process. That is, we needed to show how our proposed research fundamentally differed from the medical model by clearly articulating how a project such as Cat Royale could potentially benefit both humans and animals in different ways, without using animals as research instruments but rather involving them as research participants, who would be protected to the same extent as human participants might be protected in medical research~\cite{Mancini2022relevance}.  

\subsection{Wider reflections on risks and benefits}
 We reflect that, in negotiating these three tensions, it often appeared easier to identify and mitigate potential harms than it was to articulate benefits. The ethical process identified many risks of harm including stress and injury to the cats, reputational risks to the University and individual researchers being exposed to potentially hostile social media. Once clearly identified, however, these were then possible to mitigate (at least given appropriate external advice), for example, by planning a wide range of measures to ensure the comfort and safety of the cats, or by delegating media relations to the artists and senior members of the research team who had undertaken media training. In contrast, articulating potential benefits to balance potential harms proved far more difficult, as evidenced by the extensive dialogue with both CARE and AWERB about there being no clear experimental design, insufficient clarity over data collection and a lack of reusable results. Such discussions relate to the assumed validity of the research, and the knowledge developed through it, and ultimately speak to whether the research might deliver valuable benefits or might even be seen as dangerously bogus.  

The difficulty of demonstrating the benefits of using animals in research relative to the ease of identifying harms is well-known and has been highlighted by policy advisory working groups, such as the American Association for Laboratory Animal Science–Federation of European Laboratory Animal Science Associations, in short AALAS–FELASA~\cite{Laber:2016:Reco}. This is because, in much animal research (e.g. medical research, toxicology studies), the harms inflicted on animals are both certain and immediate, whilst the benefits to society are uncertain and may only manifest in the long term (especially in the case of fundamental research). Thus, to justify the use of animals, researchers must demonstrate that the work they propose to do can yield valid knowledge and, as discussed previously, scientific knowledge is regarded as the gold standard. In contrast, in research that conforms to animal-centred principles, as in ACI, knowingly harming animals is simply not permitted, regardless of what kind of knowledge a study might yield. This does not necessarily mean that animals might not be harmed during a project, but any harms would occur only if researchers unintentionally failed to predict and mitigate their possible occurrence. At the same time, other forms of knowledge might be regarded as valid and worth negligent risks to the animals involved, provided any risks were properly assessed and duly mitigated. However, what might be argued as benefits from an ACI perspective, might be regarded as insufficient from a scientific perspective. In other words, being able to claim benefits, and thus what risks are worth taking to deliver those benefits, is fundamentally dependent on specific epistemological perspectives and on related perceptions of what constitutes valid and legitimate knowledge. Without explicitly acknowledging, addressing and resolving epistemological tensions arising within a research project, it may be difficult to convince ethical IRBs of sufficient benefits, even when risks can be anticipated and mitigated.

\hl{A second reflection concerns reputational risk. Our IRBs explicitly raised this alongside ethical considerations of the welfare of the cats and human audience. In contrast to previous arguments that reputational risk can be a ‘serendipitous device’ to constrain academic freedom} ~\cite{librett2010apples}, \hl{our experience was that our IRBs were not seeking to close down the project, but rather were helping us anticipate and manage reputational risks to all concerned, including ourselves, while also requesting that the University be kept in the picture. Discussions of reputational risk sensitised us to potential harms to our team members of unwitting exposure to public backlash (a particular sensitivity for animal research) and enabled us to anticipate various possibilities when drawing up our communications strategy. We also came to appreciate the risk that a naïve engagement with the complexities of animal research might potentially impact on our colleagues’ freedom to conduct their own research, both in the veterinary science and autonomous systems communities. We therefore concur with  Stark’s call for greater dialogue and mutual understanding}~\cite{stark2007victims}. \hl{In general, we found the face-to-face discussions with IRBs to be extremely helpful, though recognise that they need to be carefully managed in a positive spirit. We also agree with Martin and Inwood in calling for more open recognition of the subjectivities involved~\cite{martin2012subjectivity}, and offer beneficiary-epistemology space as a mechanism for making them visible, especially when multiple IRBs are involved.}

\hl{
Finally, we reflect on whether any ethical perspectives were missing from the process.
One possibility is a perspective from the arts and humanities that might have informed discussions of artistic and aesthetic concerns.  
Such matters were not within the scope of AWERB and CARE, and while CS-REC did have experience of considering art as a method in HCI, its expertise did not cover aesthetic matters.
Stronger arts and humanities expertise might have surfaced ethical issues that were not to the foreground in our interactions with the three IRBs.
A notable example was the \textit{artistic reasoning} for choosing to work with cats as noted earlier.
Looking back, our ethics paperwork focused on the practical reasons for choosing cats, but did not speak to our artistic motivation; that their perceived combination of cuteness and autonomy might set up an uncomfortable tension in the audience.
We did not offer this artistic rationale to the IRBs, thinking it was outside their scope or perhaps fearing it would add confusion to an already complicated discussion.
Understandably, they did not raise it themselves.
Yet, in hindsight, it perhaps sat in the background as an uncomfortable feeling for our IRBs (as well as for the audience) that coloured their reaction to the project.
Introducing additional arts and humanities expertise might have helped clarify such aesthetic choices.}

\section{Implications for Future Ethical Review Process}
 We now consider how future projects could benefit from our reflections on Cat Royale and the idea of beneficiary-epistemology space. 

\subsection{Supporting explicit ethical framing of projects from the outset}
The most immediate use of beneficiary-epistemology space could be for researchers to carefully position their own HCI/ACI projects alongside their relevant local ethical IRBs. ACI projects that consider animals as their intended beneficiaries could also consider risks and benefits to humans who are likely to be implicated (e.g., owners, animal professionals, strangers who encounter them, the researchers themselves). HCI projects that intend to benefit humans could also consider risks and benefits to animals wherever they live, whether within homes, on farms, or in the wild. Many organisations will likely have specific ethical and regulatory requirements for research involving animals and complying with these will often involve dealing with IRBs whose primary remit lies outside of HCI/ACI and who may not even be familiar with these fields. 

Locating themselves, their local IRBs, their methods and wider contributing disciplines within beneficiary-epistemology space may help projects identify underlying tensions that are likely to emerge during the ethical review process when the project may be reviewed by assessors from different disciplinary backgrounds. A key lesson from our experience is that researchers should consider how they are appropriating knowledge and methods from other disciplines and whether this poses risks to researchers from these wider communities. Conversely, they might consider any potential benefits to these communities and what kinds of evidence would be seen as being convincing to them.

\subsection{Securing expertise including in Project Ethical Advisory Panels}

A second strategy for improving the ethical review process, reflecting calls for greater dialogue with IRBs~\cite{Munteanu:2015:SitEthic, stark2007victims}, is to involve experts who can help transfer knowledge, negotiate differences and build confidence, by bringing their specialist expertise into the project and bridging the different disciplines involved, using beneficiary-epistemology space to facilitate dialogue. Each project might establish a dedicated Ethical Advisory Panel at the beginning of the research design process comprising stakeholders, whose academic expertise might inform the formulation of proposed research questions and methods; these might include, for example, computer scientists, designers, artists, and animal researchers. Such panels might also involve stakeholders who represent the intended beneficiaries of the proposed research or who are likely to be affected by it (as did the Audience Advisory Panel for Cat Royale); these might include, for example, members of the public, animal guardians, animal welfare organisations, and possibly also animal justice advocates (animal representatives whose role would be to ensure that issues of animal justice are also considered~\cite{Mancini2022relevance}). Additionally, such boards might involve members of the relevant institutional ethical IRBs, who would be willing to participate in the research design process, thus representing the perspective of the relevant disciplines (formalising the kind of advice we informally received in Cat Royale). To avoid conflicts of interest, any members of ethical IRBs who participate in the research design process for a project would then be excluded from its ethical review process.  

\subsection{Establishing HCI-ACI Ethical IRBs}
The third longer-term strategy we propose is to reshape IRBs. Ideally, given a growth in demand, research institutions would establish dedicated \textit{HCI-ACI Ethical IRBs} to assess projects at the intersection of HCI and ACI, which could perhaps include selected members of existing ethical IRBs and which would cover multidisciplinary epistemological, methodological and ethical perspectives. HCI-ACI ERB members would be familiar with both HCI’s and ACI’s epistemological bases, their ethical values and their consequent methodological demands, and how these require the protection of both humans and animals from any harm, thus providing a basis for assessing and managing risks.  On the one hand, awareness of the value that the welfare and autonomy of both humans and animals involved in research must be protected would provide the bounds within which methodological flexibility can be exercised with minimal risks. On the other hand, appreciation of the value of different epistemologies would open the methodological horizon to the diverse benefits that integrating different knowledges could yield.   
\section{Conclusions}
Designing multispecies computing systems requires the integration of multidisciplinary expertise, including in the process of ethical review. This means that researchers need to be open to the epistemological and ethical perspectives of diverse disciplines, including those with which HCI and ACI’s ethical values may be in tension. As a complex project to design a computing system to benefit cats while also provoking humans to reflect on wider questions of trust in AI, Cat Royale undertook a complex journey through ethical review that involved constructive dialogue with three distinct IRBs. With hindsight, this was a powerful probe for exploring the diverse perspectives involved and consequent tensions that needed to be negotiated. We found the framework of \textit{beneficiary-epistemology} space to be useful for revealing and better understanding our own internal and external tensions, and suggest that it could guide other projects and IRBs in the future as they negotiate the complexities of ACI research. \hl{We also recognise the possibility of extending our framework to consider the ethical involvement of other non-human stakeholders beyond animals, for example, considering plants and wider environmental impacts, which might also require ethical discussions with other disciplines. It is interesting to speculate whether, in some long term future, Artificial Intelligences might ever be warranted ethical consideration as non-human stakeholders.}
For the present, we close with some final reflections that look beyond our particular focus on multispecies interactions to HCI more widely. 

We wholeheartedly agree with those who have called for greater reflection on and discussion of ethical process within our field. Our experience reveals that it can be an extremely complex matter, especially for interdisciplinary projects. We agree with~\citet{Munteanu:2015:SitEthic},~\citet{Brown:2016:FiveProvo}, and~\citet{Benford:2015:Ethical}, that HCI’s ethical process needs to be more responsive and dynamic, involving ongoing dialogue with IRBs, greater researcher responsibility on the ground, and doing ethics ‘on the way out’, rather than signing off a rigid protocol at the start. However, the process will be improved by researchers having a map to guide them along their way; this is where we suggest that leveraging \textit{beneficiary-epistemology} space can contribute.

We also sympathise with those who, from a humanities and social sciences perspective, warn about the dangers of HCI (and other fields) simply inheriting the medical model of ethics, given that this is so well established within institutions. However, it is important to bear in mind that HCI is a broad interdisciplinary endeavour engaging with diverse communities, some of which do require tight legal regulation of research (as with animal-research).  

We further emphasise that engaging with other disciplines is not a one-way street. There may be a danger to HCI or ACI researchers in being seen to appropriate other disciplines’ ideas and methods while disregarding aspects of their epistemological values and ethical processes. Similarly, there may be a danger to those communities in being seen as allowing their disciplinary standards to be compromised. If we wish to involve other researchers into our projects, then we need to account for their ethical concerns and underlying epistemologies. If we do not, we risk becoming ‘research tourists’!  

Finally, we wonder about the ethics of such ethical reflections. We did not set out to write a paper on ethical review process at the start of our project and our ethical discussions with IRBs did not cover this eventuality. We did consult with our IRBs who saw various drafts of this paper. However, we feel a need for greater discussion within HCI and with IRBs and other stakeholders regarding how we can best undertake to reflect on ethical process in a way that is beneficial to all. 

\begin{acks}
We gratefully acknowledge support from the UKRI Trustworthy Autonomous Systems Hub (Grant EP/V00784X/1) and the EPSRC Centre for Doctoral Training in Horizon: Creating Our Lives in Data (Grant EP/S023305/1). We are also very grateful for the time taken by the AWERB, CSREC and CARE IRBs to review and shape our project, and to the many others who contributed including Professor Daniel Mills of Lincoln University, the Royal Society for the Prevention of Cruelty to Animals (RSPCA), and the members of our Audience Advisory Panel. And of course, special thanks also to Clover, Pumpkin, Ghostbuster, and Ed.
\end{acks}

\bibliographystyle{ACM-Reference-Format}
\bibliography{ref}

\end{document}